
\magnification=\magstep1
\centerline{\bf ON THE ROLE OF INITIAL DATA IN}
\centerline{\bf GRAVITATIONAL COLLAPSE OF INHOMOGENEOUS DUST}
\vskip 1 in
\centerline{P. S. Joshi and T. P. Singh}
\centerline{Theoretical Astrophysics Group}
\centerline{Tata Institute of Fundamental Research}
\centerline{Homi Bhabha Road, Bombay 400 005}
\centerline{INDIA}
\vskip 1 in

\magnification=\magstep1  
\hoffset=0 true cm        
\hsize=6.0 true in        
\vsize=8.5 true in        
\baselineskip=24 true pt plus 0.1 pt minus 0.1 pt
\overfullrule=0pt         

\centerline{\bf ABSTRACT}
\smallskip
\noindent We consider here the gravitational collapse of a spherically
symmetric inhomogeneous dust cloud described by the Tolman-Bondi models.
By studying a general class of these models, we find that the end state of the
collapse is either a black hole or a naked singularity, depending on
the parameters of the initial density distribution, which are $\rho_{c}$,
the initial central density of the massive body, and $R_0$, the initial
boundary. The collapse ends in a black hole if the dimensionless
quantity $\beta$ constructed out of this initial data is greater than 0.0113,
and it ends in a naked singularity if $\beta$ is less than this number.
A simple interpretation of this result can be given in terms of the strength
of the gravitational potential at the starting epoch of the collapse.

\vfil\eject

\noindent The gravitational collapse of a sufficiently massive
homogeneous dust ball
leads to the formation of a black hole, as was indicated by the work of
Oppenheimer and Snyder [1], and subsequent investigations.
Such a black hole covers the infinite density
singularity forming at the center of the cloud, which is the end state
of collapse. However, the final fate of collapse of an inhomogeneous
distribution of matter with a general equation of state is largely unknown.
An important subcase which could be examined in this context is the
spherically symmetric but inhomogeneous
distribution of dust which collapses under the force of gravity. The general
solutions to Einstein equations for this case have
been given by the Tolman-Bondi spacetimes [2], and it has been
demonstrated that naked singularities can occur as the end state of such
a collapse [3, 4, 5, 6].
In particular, it was pointed out in [5, 6] that the collapse ends in a naked
singularity or a black hole depending on whether or not a certain
algebraic equation involving the
metric functions and their derivatives has positive real roots.

In order to be able to
ascertain the astrophysical implications of such a result, it is
necessary  to
translate the condition for the existence of positive real roots into
the actual constraints
on the initial density distribution in the cloud.

We investigate this issue in the present paper for a class of models of the
Tolman-Bondi spacetimes. One would expect
in this case that the degree of inhomogeneity in the matter distribution
plays a role in determining the final fate of the collapse. This is
demonstrated here by working out the explicit conditions for the
collapse to the endstate which is either a black hole or a naked
singularity depending on the initial conditions chosen. It turns out that
for the class we are considering, these outcomes are characterized in terms
of the existence of real positive roots of a quartic equation
which we shall determine here.
This enables us to relate the black hole or naked singularity
configuration as the end state of the gravitational collapse in terms of the
initial density distribution $\rho_c$ and radius $R_0$ of the massive
body. Finally, we demonstrate our procedure by an explicit example
of a typical initial density profile from which the collapse develops.

We use the comoving coordinates $(t,r,\theta,\phi)$ to describe
the spherically symmetric collapse of an inhomogeneous dust
cloud. The coordinate $r$ has non-negative values and
labels the spherical shells of dust and $t$ is the proper
time along the world lines of particles given by $r=$
constant. The collapse of spherical inhomogeneous
dust is described by the Tolman-Bondi metric in comoving
coordinates (i.e. $u^i=\delta^i_t)$:
$$ds^2=-dt^2+
{R^{\prime 2}\over
1+f} dr^2+R^2(t,r)(d\theta^2+\sin^2\theta d\phi^2).\eqno(1)$$
The energy momentum tensor is
$T^{ij}=\epsilon\delta^i_t\delta^j_t$, where $\epsilon$ is the energy
density. From the Einstein equations it follows that
$$\epsilon=\epsilon(r,t)=
{F^{\prime}\over R^2R^{\prime}},\qquad
\dot R^2={F\over R}+ f.\eqno(2)$$
(We have set $8\pi G/c^{4}=1$). Here the dot and prime denote partial
derivatives with respect to $t$ and $r$ respectively.
The quantity $F(r)$ arises as a free function from the integration
of the Einstein equations and can be interpreted physically as the total
mass of the collapsing cloud within a coordinate radius $r$. Thus we take
$F\ge0$ for the mass function $F(r)$.
The quantity $R(t_1,r_1)$ denotes the physical radius of a shell of
collapsing matter at a coordinate radius $r_1$ and on the time slice $t=t_1$.
The quantities $F$ and $f$ are arbitrary functions of $r$. In further
discussion, we restrict to the class of solutions
$f(r)\equiv 0$, which are the marginally bound Tolman-Bondi models.
Similar considerations could, however, be developed for the models with
$f>0$ or $f<0$. As we are concerned with the collapsing cloud, we
take $\dot R(t,r)<0$.

The epoch $R=0$ denotes a physical
singularity where the spherical shells of matter collapse to zero radius and
where the density blows up to infinity. The time $t=t_0(r)$ corresponds to
the value $R=0$ where the area of the shell of matter at a constant
value of coordinate $r$ vanishes. This singularity curve $t=t_0(r)$
corresponds to the time when the matter shells meet
the physical singularity. This specifies the range of coordinates for
the metric (1) as
$$ 0\le r<\infty, \quad  -\infty< t< t_0(r),$$
whereas $\theta$ and $\phi$ have the usual coordinate range. In the case
of a finite cloud of dust, there will be a cut off at $r=r_b$, where the
metric is matched smoothly with a Schwarzschild exterior.

The Tolman-Bondi models admit a freedom of scaling in the following
sense. One could arbitrarily relabel the dust shells given by
$r=$const. on a given $t=$const. epoch,
by letting $r\to g(r)$. Thus, at any constant time surface,
say at $t=t_0$, $R(r,t_0)$ can be chosen to
be an arbitrary function of $r$.
This arbitrariness  reflects essentially the
freedom in the choice of units.
For convenience of calculation, we make the choice of scaling
at $t=0$ as given by $R(r,0)=r$.
With this scaling, the $\dot{R}$ equation in (2), (with $f=0)$, can be
integrated to get
$$R^{3/2}(r,t)=r^{3/2}-{3\over 2}\sqrt{F(r)}t,\eqno(3)$$
and the energy equation becomes
$$\epsilon(r,t)=
{4/3\over
\left(t-{2\over 3}{G(r)\over H(r)}\right)
\left(t-{2\over 3} {G^{\prime}(r)\over H^{\prime}(r)}\right)},\eqno(4)$$
where $G(r)=r^{3/2}$,
$G^{\prime}(r)=(3/2)r^{1/2}$, and $H(r)= \sqrt{F(r)}$.
We now write
$F(r)\equiv r\lambda (r)$ and assume
$\lambda(0)\equiv \lambda_{0}\not= 0$ and finite, which is the class of models
considered in Ref. 5. This means that near the
origin, $F(r)$ goes as $r$ in the present scaling, and
the density at the center behaves with time as
$\epsilon(0,t)={4/ 3 t^2}$. This is a general class of models which
includes all self-similar solutions as well as a wide range of
non-self-similar models, which we find quite adequate for the purpose
of present investigation.
The central density becomes singular at
$t=0$,
and the singularity is
interpreted as having arisen from the evolution of dust collapse
which had a finite density distribution in the past on an
earlier non-singular initial epoch.

To check if the
singularity could be naked, what needs to be examined is
the possibility that future directed
null geodesics could come out of the singularity
at $t=0$, $r=0$. The equations of outgoing
radial null geodesics in the
space-time, with $k$ as affine parameter, can be written as,
$${dK^{t}\over dk} + \dot R^{\prime}
K^rK^t=0,\eqno(5)$$
$${dt\over dr}={K^{t}\over K^r}=R^{\prime},\eqno(6)$$
where $K^{t}=dt/dk$ and $K^r=dr/dk$ are tangents
to the outgoing null geodesics.
The partial derivatives $R'$ and $\dot{R}'$ which occur in (5) and (6)
can be worked out from Eqn. (3) and are most suitably written as
follows,
$$R^{\prime}=\eta P-
\left[{1-\eta\over \sqrt{\lambda}}+
\eta {t\over r}\right]\dot R\eqno(7)$$
$$\dot R^{\prime}=
{\lambda\over 2r P^2}
\left[{1-\eta\over \sqrt{\lambda}}+\eta
{t\over r}\right]\eqno(8)$$
where we have put
$$ R(r,t)=r P(r,t),\; \eta=\eta(r)={rF^{\prime}\over F}.\eqno(9)$$
The functions $\eta (r)$ and $P(r)$ have been introduced since they
have a well-defined limit in the approach to the singularity.

If the outgoing null geodesics terminate in the past with a
definite tangent at the singularity (in which case the
singularity would be naked), then using (6) and l$^{\prime}$
Hospital rule we get
$$X_0=\lim_{t\to 0,r\to 0}
{t\over r}=
\lim_{t\to 0,r\to 0}
{dt\over dr}=\lim_{t=0,r=0}
R^{\prime}\eqno(10)$$
where $X=t/r$ is a new variable. The positive
function $P(r,t)=P(X,r)$ is then given
using (3) by
$$X-{2\over 3\sqrt{\lambda}}=
- {2P^{3/2}\over 3\sqrt{\lambda}}\eqno(11)$$
We define $Q=Q(X)=P(X,0)$.
If future directed null geodesics come out of the singularity at
$t=0,r=0$, meeting the singularity in the past with a definite tangent
$X=X_0$ as given above, then it follows from (11) that such a value $X_0$
must satisfy
$$ X_0<  {2\over 3\sqrt{ \lambda_0}}$$

Further, we note, by using the
definition $F=r\lambda(r)$, that as $r\rightarrow 0$,
$\eta\rightarrow 1$. Also, from (2) it follows that for
$f=0$, $\lim \dot R = -\sqrt{\lambda_{0}/Q}$. Using these results
in the expression (7) for $R'$ implies that the condition (10) is
simplified to the equation
$$V(X_{0})=0\eqno(12)$$
where
$$V(X)\equiv Q+X\sqrt{{\lambda_0\over Q}}-X.$$

In order to be the past end-point of outgoing
null geodesics, at least one real positive value of $X_{0}$ must satisfy (12)
(of course, subject to the constraint
implied by (11) as stated above, i.e. $X_0< 2/3\sqrt{\lambda_0}$).
In general, it is seen [5] that if the equation
$V(X_{0})=0$ has a real positive root, the singularity would be naked.
Whenever this is not realized, the evolution will lead to a black hole.
Such a singularity could be either locally or globally naked depending on
the global features of the function $\lambda(r)$.

We should clarify the sense in which the terms naked singularity
and black hole are used. When there are no
positive real roots to equation (12), the central singularity is not naked,
because it follows that there are no outgoing future directed null geodesics
from the singularity in that case. Further, it
is known that the shell focusing singularity $R=0$ for $r >0$
is always covered (for a proof see, e.g. [4], [7]). Hence in the absence
of positive real roots the collapse will always lead to a black hole.
On the other hand, if there are positive real roots, it follows that the
singularity
is at least locally naked, though for brevity we have simply called it a
naked singularity throughout the paper. Such a locally naked singularity
would be globally naked as well when the outgoing trajectories could reach
arbitrarily large values of $r$ (i.e. the signals reach far away observers).
Otherwise there would still be a black hole when these trajectories fall
back to the singularity without coming out of the horizon. This is a
violation
of weak censorship only. The occurrence of either of these situations will
depend actually on the nature of the function $\lambda(r)$.
The conditions under which this locally
naked singularity could be globally naked as well have been discussed,
for instance, in [6], and we do not go into them here. The occurence
of positive real roots implies the violation of strong cosmic censorship,
though not necessarily of weak cosmic censorship. In other words, black
hole and locally naked singularity are not mutually exclusive alternatives.
It can also be shown that whenever there is a positive real root to (12), a
family of outgoing null geodesics always terminates at the singularity in the
past [6].

We now examine the condition for the occurrence of a naked singularity
in some detail. Using (11), the condition $V(X_{0})=0$ can be written as
$$ Y^{3}(Y-{2\over 3}) - \alpha (Y-2)^{3}=0         \eqno(13)$$
where we have set $Y=\sqrt{\lambda_{0}} X_{0}$,
$\alpha=\lambda_{0}^{3/2}/12$. (Recall that $F(r)$ and hence $\lambda_{0}$
are positive). Using standard results it can be
shown that this quartic equation has positive real roots if and only
if $\alpha>\alpha_{1}$ or $\alpha<\alpha_{2}$, where
$$\alpha_{1}={26\over 3} + 5\sqrt{3}\approx 17.3269$$
and
$$\alpha_{2}={26\over 3} -  5\sqrt{3}\approx 6.4126\times 10^{-3}.$$
To derive this condition,
first note from (13) that if it has a real root, it must be
positive as negative values of $Y$ do not solve this equation.
Writing the general quartic as
$ ax^{4} + 4bx^{3} + 6cx^{2} + 4dx + e =0$
one defines $H=ac-b^{2}, I=ae-4bd+3c^{2},
 J=ace+2bcd-ad^{2}-eb^{2}-c^{3}$ and $\Delta = I^{3} - 27 J^{2}$.
If $\Delta< 0$ the quartic has two real and two imaginary roots.
If $\Delta> 0$, all roots are imaginary unless $H< 0$ and
$(a^{2}I-12H^{2})< 0$, in which case they are all real. The application of
this procedure to the
quartic in (13) leads to the condition on $\alpha$ given above.

It follows, however, from (11) as stated earlier that along any such
outgoing null geodesics we must have $\sqrt{\lambda_0} X_0= Y<2/3$. Then
(13) implies that the larger range of $\alpha$ for the existence of roots,
i.e. $\alpha> 17.3269$
is ruled out in the sense that no outgoing trajectories can meet the
singularity with this larger value of tangent $X_0$.
This is seen from (13) by writing $\alpha$ as a function of $Y$,
which tells that if $\alpha>\alpha_{1}$, then $Y>2$.
It thus follows that a naked singularity arises if and only if
$\alpha<\alpha_{1}$, or equivalently, if and only if $\lambda_{0}<0.1809$.
Whenever the limiting value $\lambda_0$
does not satisfy this constraint the gravitational collapse of the
dust cloud must end in a black hole. The physical interpretation for
the quantity $\lambda_{0}$ can be obtained from
Eqn.(4) for the time evolution of the energy density. If the collapse
starts at a time $-t_{0}< 0$, and $\rho_{c}$ is the initial
energy density at the center, then $\rho_{c}=4/3t_{0}^{2}$.
If $\rho_{c}'$ be the initial density gradient at the center,
then we find from (4) that
$\lambda_{0} = 16{\rho_{c}^{3}/3\rho _{c}^{\prime 2}}$.
Putting in the units gives that
$$\lambda_{0}=  {8\pi G\over c^{4}} {16\rho_{c}^{3}\over 3\rho _{c}
^{\prime 2}}\eqno (14)$$
Defining $\beta=\lambda_0/16$, we find that the black hole arises whenever
$$0.0113<\beta\eqno(15)$$
and the naked singularity results for the values given by
$\beta<0.0113$.
The occurrence of one or the other is governed by conditions on
a combination of the initial central density and the initial density
gradient at the center. The cosmic censorship hypothesis of Penrose [8]
could, in the present context, be translated to the conjecture that
values of $\beta$ smaller than 0.0113 do not occur in realistic collapse.

One needs to calculate now the value of the parameter $\beta$ or
$\lambda_0$, given the initial density profile for the collapsing massive
star. Given the initial central density, one can evaluate the initial
density gradient $d\rho/dr =\rho_{c}'$ at the center as follows.
Firstly, note that using (4) one can write the expression for
$d\rho/dr$ and it is seen that in the limit of $r\to 0$ this always goes
to a finite quantity which is proportional to $1/\sqrt{\lambda_0}$. Now, given
the initial data in the form of the density distribution $\rho(R)$ for
the body at an initial non-singular epoch of time, in terms of the
physical radius $R$, one can integrate using (2) to get the mass function
$F(R)$. Then (3) provides a functional relationship $r(R)$ which can
be inverted (in principle) to express $R$ in terms of $r$. One could then
write the  mass profile explicitly as $F(r)$ and $\lambda_0$ is evaluated as
the limit of $F(r)/r$ as $r\to 0$.

These results can be highlighted with an explicit example. To check whether
the black hole or naked singularity could occur, one will start by
examining the initial density profile or mass profile for the collapsing
body and will go on to calculate $\lambda_{0}$ as pointed out above.
Consider, for instance, the following initial
density profile, given at a non-singular epoch $t=-t_{0}$, as a
function of the physical radius $R$:
$$    \epsilon (R,-t_{0}) \equiv \rho (R) =
     \cases{ \rho_{c} (1 - {R^{3}/ R_{0}^{3} }), & $R\le R_{0}$\cr
                    0, &  $R > R_{0}$\cr }
                                                 \eqno (16)   $$
$\rho_{c}$ is the central density and $R_{0}$ the initial boundary of the
object.
Using (2), it follows from the equation for $\epsilon (r,t)$ at a constant
time epoch that $F(r)$ can be written as a function of R as
$$    F(R) = {1\over 3} \rho_{c} R^{3}
        \left ( 1 - {R^{3}\over 2R_{0}^{3}}\right ).
                                            \eqno (17) $$
The relation between $R$ at the initial epoch $-t_{0}$ and $r$ can be found
from Eqn. (3) using the above form of $F(R)$. We get
$$      r^{3/2}   =   R^{3/2}
\left [ 1 - \left( 1 - {R^{3}\over 2 R_{0}^{3}}\right)^{1/2}\right ]
                                                             \eqno (18)$$
We are interested in finding the form of $F(r)$ near the center, so in
(18) we take $R\ll R_{0}$ and find after a binomial expansion that
$R^{3} = (16)^{1/3} R_{0}^{2} r$ near the center.  Using this in (17)
shows that near $r=0$, the form of $F(r)$ is
$$   F(r) =  {1\over 3} (16)^{1/3} \rho_{c} R_{0}^{2} r
             \left [ 1 - {(16)^{1/3}\over 2} {r\over R_{0}}\right ].
                                           \eqno (19)    $$
Hence, $\lambda_{0}$, which is the limiting value of $\lambda (r)=F(r)/r$, is
given by
$\lambda_{0} = (16)^{1/3} \rho_{c} R_{0}^{2}/3$. We see that $\lambda_{0}$
is defined in terms of parameters of the original density profile
given by Eqn. (16), as expected. After putting in the units, and assuming
$\rho/c^{2}$ to be the mass density of the collapsing object, and using
the condition (15) on $\beta$, we find that for the present density
profile, given by (16), a black hole will form whenever
$\rho_{c}R_{0}^{2}/c^{2}$ lies in the range
$$   1.16\times 10^{26} gms/cm\ < \
\rho_{c}R_{0}^{2}/c^{2}  \eqno(20)$$
For a comparison, we note that for a neutron star with a central density of
$10^{15}gms/cm^{3}$ and radius $10^{6}$ cm,
$\rho_{c}R_{0}^{2}/c^{2} = 10^{27} gms/cm$.

We can calculate here the initial density gradient $d\rho/dr=\rho_{c}'$
at the center, using the form (16) of the density
profile, and the relation between $r$ and $R$ near $r=0$. We
get $\rho_{c}'=-(16)^{1/3} \rho_{c}/R_{0}$ which coincides with (14) once
we express $R_{0}$ in terms of $\lambda_{0}$ and $\rho_{c}$.
Note that in this example, while the physical density gradient
$d\rho/dR$ is zero at the center, the coordinate gradient $d\rho/dr$
at the center is non-zero, because the derivative $dr/dR$ at the fixed
initial epoch goes to zero at $r=0$. In fact, we can draw some general
conclusions about the density profiles considered in this paper.
As indicated earlier, $\rho_{c}'$ goes as $1/\sqrt{\lambda_{0}}$, and is
hence non-zero. Using (3), we can write, near $r=0$,
$${d\rho\over dr} = {d\rho/dR \over dr/dR} =
       {3 \lambda_{0} t_{0}^{2}d\rho/dR\over 4 R^{2} } \eqno (21)$$
Since $\rho_{c}'$ is non-zero and finite, it follows that $d\rho/dR$
should behave as $R^{2}$ near the center. In other words, $d\rho/dR$
and $d^{2}\rho/dR^{2}$ should be zero at $r=0$, and $d^{3}\rho/dR^{3}$
should be non-zero.

Let us define $\rho_{3}$ to be the value of $d^{3}\rho/dR^{3}$ at
the origin. It is then straightforward to check for the density
profile in (16), as well as for the general density profiles considered
in this paper, that $\beta$ is proportional to
$\rho_{c}^{5/3}/\rho_{3}^{2/3}$. Hence it is apparent that suitable
density gradients, (as measured by $\rho_{3}$), will lead,
through the dynamical time evolution of the collapse, to a naked
singularity. On the other hand, when $\rho_{3}$ is strictly zero,
the model is like the Oppenheimer-Snyder model for which the singularity
is covered. We note, however, that it need not be the same as
the Oppenheimer-Snyder model, if the higher order derivatives
are non-zero.
Thus we have the situation that while homogeneous collapse
leads to a black hole, a suitable amount of inhomogeneity, as
represented by a non-zero $\rho_{3}$, leads to a naked singularity.
One should note that departure
from the homogeneous models by way of introducing arbitrarily small
inhomogeneity
does not necessarily make the singularity naked, as pointed out
by our calculations here. This agrees with the earlier results such as
those in [3,5,6].

Our analysis so far has been exact, within the framework of
Tolman-Bondi models. While these models may not be sufficiently
complex as to describe the final fate of real gravitational collapse,
they can be useful for the purpose of obtaining insights. With this in
mind, we consider the following Newtonian order of magnitude estimates.
Even though $\rho_{c}'$ is the central density gradient, defined using the
coordinate radius $r$, we approximate it to be the mean initial density
gradient across the collapsing star. If for an object like a neutron
star, we assume the linear extent to be $L\approx 10^{6} cm$, and
assume $\rho_{c}$ to be ($c^{2}\times$ mass density), and take
$\rho_{c}/c^{2}\approx 10^{15}gms/cm^{3}$, and
$\rho'_{c}\approx \rho_{c}/L$, we
get $\beta=0.62$. We note that the model yields
a value of $\beta$ in the range given in Eqn. (15), for which the collapse will
lead to a black hole, if no force other than gravity is acting.
We should emphasize that while the condition for the occurrence
of the naked singularity is exact when stated in terms of the quantity
$\beta$, the argument given above is only an approximation
and is included here as it seems to suggest that the bounds on
$\beta$ given here may have relevance to real collapse.

The dimensionless quantity $\beta$ is essentially
$2GM\gamma/Lc^{2}$, where
$M= 4\pi L^{3}\rho_{m}/3$ is approximately the mass of the collapsing body
having an initial linear extent $L$
and an initial mean density $\rho_{m}$, and $\gamma=\rho_{c}/\rho_{m}$.
Looked at in this way, $\beta/\gamma$ is nothing
but an estimate for the gravitational potential of the body, expressed
in the basic potential units $c^{2}/ G$. It then appears natural that
$\beta$ should decide whether the collapse ends in a black hole or
a naked singularity. Once $\beta$ becomes sufficiently large, the collapse
appears to proceed to a black hole. From an astrophysical viewpoint,
it is of interest to investigate whether values of $\beta$ large enough
to lead to a black hole can always be realized in realistic collapse.
One also notes that $\beta$ is  inversely proportional to the degree
of inhomogeneity at the center. (This follows from (14)).
The more inhomogeneous the system, the smaller is $\beta$.

This model can be generalized in many ways, for instance by considering
the most general class of functions $F(r)$ and $f(r)$.
 Also, it has
been shown [9] that the pattern of a transition from the black-hole
configuration
to the naked singularity configuration persists in models with more
general equations of state. It is desirable to cast results for these
models in terms of constraints on the initial density distribution.
Work is in progress to make this connection rigorous.
\medskip

Acknowledgements: We thank I. H. Dwivedi, S. M. Chitre and
Alak Ray for helpful comments.  We also thank the
referees for suggesting modifications which improved the paper.

\bigskip
\centerline{\bf REFERENCES}
\smallskip
\item{[1]} J. R. Oppenheimer and H. Snyder (1939), Phys. Rev. 56, 455
\item{[2]} R. C. Tolman (1934), Proc. Nat. Acad. Sci. USA 20, 410;
           H. Bondi (1948), Mon. Not. Astron. Soc. 107, 343
\item{[3]} D. M. Eardley and L. Smarr (1979), Phys. Rev. D 19, 2239;
           D. Christodoulou (1984), Commun. Math. Phys. 93, 171
\item{[4]} R. P. A. C. Newman, Class. Quantum Grav. {\bf 3}, 527 (1986)
\item{[5]} I. H. Dwivedi and P. S. Joshi (1992), Class. Quantum Grav.
           9, L39
\item{[6]} P. S. Joshi and I. H. Dwivedi (1993), Phys. Rev. D 47,
           5357
\item{[7]} P. S. Joshi (1993), `Global aspects in gravitation and cosmology',
           Clarendon Press, Oxford; Section 6.8
\item{[8]} R. Penrose, 1969, Riv. Nuovo Cimento Soc. Ital. Fis.
           1, 252
\item{[9]} For a review on these and related developments, see for example,
           Ref. [7] above

\end